# Niépce-Bell or Turing: How to Test Odor Reproduction

David Harel
The Weizmann Institute of Science

**Abstract**

In a 1950 article in *Mind*, decades before the existence of anything resembling an artificial intelligence system, Alan Turing addressed the question of how to test whether machines can think, or in modern terminology, whether a computer claimed to exhibit intelligence indeed does so. The current paper raises the analogous issue for olfaction: how to test the validity of a system claimed to reproduce arbitrary odors artificially, in a way recognizable to humans, in face of the unavailability of a general naming method for odors. Although odor reproduction systems are still far from being viable, the question of how to test candidates thereof is claimed to be interesting and nontrivial, and a novel method is proposed. To some extent, the method is inspired by Turing's test for AI, in that it involves a human challenger and the real and artificial entities, yet it is very different: our test is conditional, requiring from the artificial no more than is required from the original, and it employs a novel method of immersion that takes advantage of the availability of near-perfect reproduction methods for sight and sound.

## 1. Background

In his famous 1950 paper in *Mind*, Alan Turing addressed the question of how to tell whether machines can think, or, using contemporary terminology, whether a computer has achieved human-like intelligence. As is well known, Turing proposed an imitation game for this, better known as the Turing test. Despite the lively controversy that has grown around his actual testing method, Turing's work is considered to have had a tremendous impact on the later research field of artificial intelligence [17]. Interestingly, the impact of Turing's paper was never diminished by the fact that at the time an intelligent computer was nowhere in sight. Perhaps to the contrary: the importance of his work is not only in the actual test he proposed but in the very raising of such a question at such an early stage, especially in view of the fact that rigorously defining intelligence was not a possibility. Needless to say, true intelligent computers in any accepted sense of the word are still nowhere in sight, even today, so many decades later, and the same goes for computers that can pass stringent versions of Turing's test. In any event, Turing's work can be viewed as one of the most interesting thought experiments in science.

In the present article, I wish to raise the not dissimilar question of how to test the validity of a candidate system for reproducing general odors artificially in a way that is recognizable to humans, in face of the fact that there are no accepted means for naming or describing odors in general. I will argue that the question itself is important and nontrivial, despite the fact that such systems are nonexistent and are far from being viable. The solution proposed here is inspired to some extent by Turing's solution to his problem, but, as we shall see, it is very different (and, of course, its potential is far more modest than testing for computerized intelligence).

In contrast to olfaction, reproduction methods for sight and sound go back to the 19th century. Nicéphore Niépce is considered to have produced the first recognizable photograph, in 1826 or 1827; see Fig. 1. The first photograph to include people was apparently the one taken by Louis Daguerre in 1838; see Fig. 2. In 1876, exactly half a century after Niépce's achievement, Alexander Graham Bell made the first telephone call, successfully summoning his assistant from the next room; see Fig. 3. In both cases, the generated artefacts were immediately recognized as being satisfactory renditions of the originals. Not perfect, of course, but unmistakably recognizable. Hence, we may say that photography, telephony and their modern offspring are, adequate reproduction methods, at least for people of compatible backgrounds.

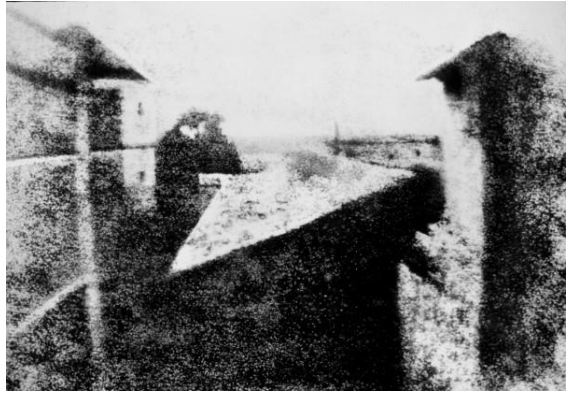

Figure 1. "*View from the Window at Le Gras*", by Nicéphore Niépce (1826–7)

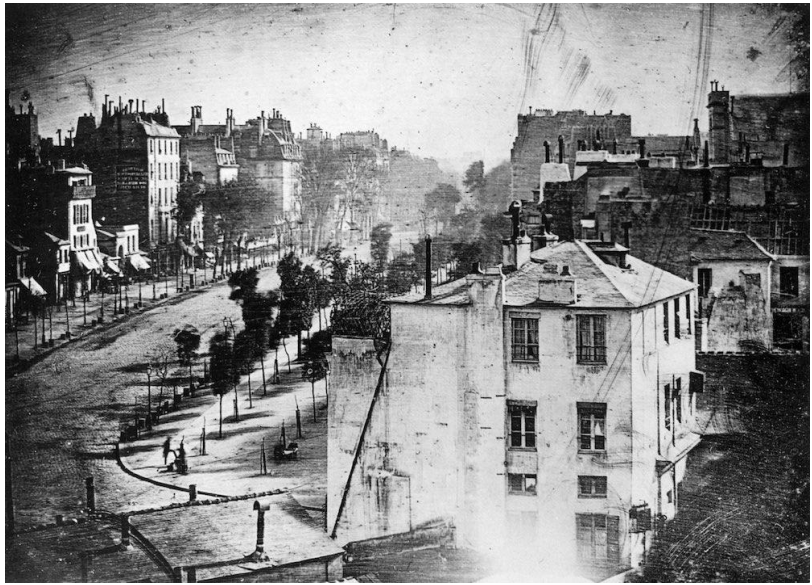

Figure 2. "*Boulevard du Temple*", by Louis Daguerre (1838)

In contrast to recognizability, perfect reproduction would call for the produced result to be indistinguishable from the original, as per Turing's test. In fact, Turing can be viewed as the best-known promoter of testing for indistinguishability. His test calls for a challenging human to try to distinguish the claimed-to-be-intelligent computer from another human by means of typed electronic communication. If, in general, the challenger is not able to tell which is which, the computer passes the test, having imitated a human's intelligent to-and-fro discussion abilities to the point of the artificial being indistinguishable from the real thing.[1] However, as discussed later, one cannot expect to achieve true indistinguishability even for the well-understood sensory modalities of sight and sound.

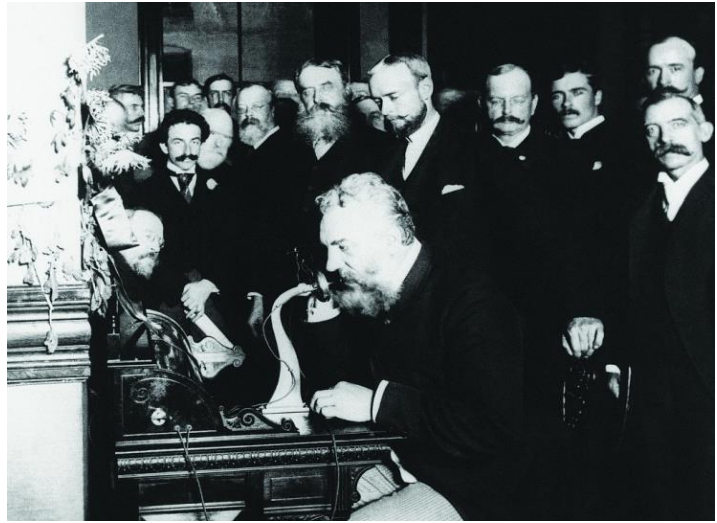

Figure 3. Alexander Graham Bell, circa 1892.

The present paper is not about intelligence; nor does it deal with sight or sound. It is about olfaction, which is the least understood of the senses, and it concentrates on the issue of testing an artificial odor reproduction system as a means for mimicking the human recognition of smell.

Olfaction has attracted a tremendous amount of deep research over the last few decades, especially since Buck and Axel's discovery of the multigene odor receptor family [1]. Parts of the work have been aimed at elucidating the neuronal mechanisms for odor processing in the brain; see, e.g., [11, 12, 14]. Others deal with developing electronic and chemical devices (e.g., gas chromatographs and specific kinds of electronic noses) for sensing and analyzing odors and for producing digital odor signatures for various applications [2, 15, 18]. Yet others deal with odor presentation; i.e., emitting odors from a pre-prepared collection, while controlling and varying concentration and flow; see, e.g., [8]. Some of the most interesting work involves the links between the various spaces relevant to olfaction; namely the chemical/molecular space, the neuronal space, the mathematical/computerizable space of electronic-nose signatures (usually vectors of numbers capturing the response of the device's sensors), and perhaps most importantly, the perceptual space (the way an odor is perceived by a human). Examples include predicting pleasantness of an odor from an e-nose signature thereof [4], linking the molecular structure of odors with human perception thereof[10], defining metrics for measuring the distance between odors and correlating it with perception [3], and attempting to predict behavioral outcome from neuronal patterns in the olfactory system [5].

---

[1] More recently, a variant of the Turing test has been proposed for checking the validity of a computerized model of a biological systems and other artefacts from nature, where an expert challenger uses probes to try to distinguish the claimed-to-be-valid model from the real thing – say, in an appropriate laboratory [6].

## 2. Odor Reproduction

Despite all of the above, what is known about the sense of smell appears to be but the tip of the iceberg. In particular, we are still very far from achieving the holy grail of the field, for which the term *artificial olfactory reproduction* seems apt: the ability to record and remotely produce recognizable renditions of arbitrary odors. We can reconstruct a visual stimulus by the spatial distribution of its electromagnetic wavelength and luminance, and for sound pitch, loudness and timbre of soundwaves in the air define a tone; all these can be readily analyzed and simulated. Odors, however, come in the form of actual molecules that our olfactory system senses, transmitting appropriate signals to the brain for perception, and little is known about the way our brains process that information and form our odor perception. Hence, analyzing and synthesizing smell is not just a question of using an appropriate set of mathematical functions to emit outputs involving accurately computed and produced wavelengths.

In direct analogy with, e.g., a digital camera and a printer, we shall consider an *odor reproduction system* (ORS) to consist of: (i) an input device, the *sniffer*, which captures and encodes certain characteristics of any input odor and transforms them into a digital signature, or fingerprint; (ii) an output device, the *whiffer*, which contains a pallete of fixed odorants with means for mixing them at high resolution and releasing the mixture to the outside world in carefully measured quantities and concentration, and with precise timing, through some appropriate aperture;[2] and, most significantly, (iii) a *mix-to-mimic algorithm*, which analyzes the signature coming from the sniffer and instructs the whiffer as to how it should mix its pallete odorants in order to produce an output odor, which, as perceived by a human, is as close as possible to the original input; see [7] and the illustration in Figure 4.

I should add that a whiffer emitting mixtures of a fixed set of odorants for the human to smell is how we see a potential future output system from today's vantage point. In the future, however, whiffers may work in totally different ways, unknown to us now; e.g., by constructing molecules on their own or by somehow triggering an appropriate brain response directly. This, however, does not affect the ideas put forward here about assessing the fidelity of such systems, which are relevant to any potential ORS, based on any kind of technology.

I shall not attempt to discuss the feasibility of constructing an artificial olfactory reproduction system. Some ideas on how that might eventually be achieved have been put forward in, e.g., [7]. Instead, I will assume that we are presented with a black-box candidate ORS, and will attempt to address the question of what it is we really want, and to seek ways to test whether we really have it. Besides these being nontrivial issues, as we shall see, I feel that they represent an intellectual challenge that is a worthwhile topic for serious contemplation.

---

[2] The quantity and concentration should be high enough to convey the sense of the odor to a human, but sufficiently low to make it possible to switch odors with negligible lingering residue.

Moreover, a serious approach to them could have practical implications for future odor-emitting and odor-reproducing systems, in a wide spectrum of application areas and in many kinds of industries.

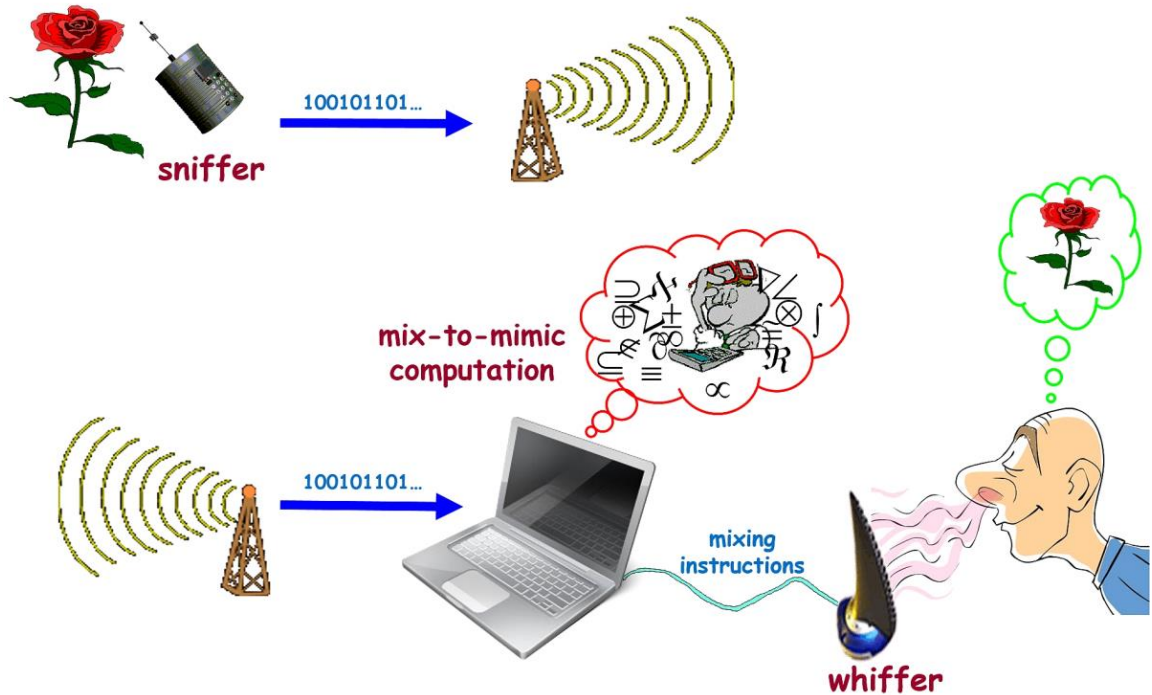

Figure 4. An odor reproduction system (ORS).

## 3. Recognition or Imitation?

So, when can we say that we have an adequate system for odor reproduction? Is it recognizability that we want, or indistinguishability? Should we opt for a Niépce-Bell-style approach (human recognizes odor in an appropriate real world sense) or a Turing-style one (human cannot tell the real odor apart from the reproduced one)? And once we decide, how should we go about setting up the appropriate test?

As a starting point, we may ask what really is involved in human recognition. We recognize and "understand" what we see in a photograph or on a video screen although we know that it is not the real thing. And the same applies to the sound coming out of a telephone or an audio system. This has been true from the very beginning. Bell's assistant in the adjacent room understood (and immediately acted upon) the famous 1876 utterance "Mr. Watson, come here – I want to see you", and no one thought of asking whether a different sentence, such as "Mr. Watson, what is the time?" would have worked too. When we view the 1826-7 and 1838 photographs, we recognize and understand what they capture and may quite safely conclude that the techniques used were general enough and could have been applied to other scenes too.[3]

---

[3] Clearly, severe technological limitations applied at the time. Niépce's photography was probably unable to

In contrast, such basic recognition of one or two photos and a couple of spoken sentences does not seem to be adequate for olfaction, for two main reasons. First, we cannot make do with trying out the system on just a few odors, especially if these are chosen by the designer of the ORS. Perhaps for a rose, an orange and a cup of coffee the system would do a good job, but not for moss in a dark cave, for screeching tires, or for the odor of an unknown animal in a faraway forest. The whiffer might simply have several particular and well-known odors built in, which it knows how to reproduce from the sniffer-generated fingerprint. Rather, we need to convince ourselves that the reproduction system works for all appropriate inputs, where the term "appropriate" pays tribute to the fact that what would be acceptable as good reproduction today would not suffice in the context of future technologies; see previous footnote.

Most severe, however, is the question of naming. How are we to become convinced that a person has "understood" an arbitrary odor and recognizes it? No methods exist for verbally describing the essence of arbitrary odors. Some attempts have been made to devise odor vocabularies, employing descriptors like musky, putrid, floral and ethereal [19]. Other work has concentrated on particular idiosyncratic fields, such as winery [13]. However, all of these appear to be deficient as general methods; they are currently not able to reliably span the entire spectrum of human-recognizable odors in a fully discriminatory fashion.

So should we opt for Turing's approach? On the face of it, a Turing-like imitation game appears to be a better bet, where a challenging person would try to invalidate the candidate ORS by distinguishing real input odors from those produced by the system. However, in Turing's test for intelligence, the human interrogator him/herself is in a deep way part of the testing process, which is not a one-dimensional "clinical" kind of test, checking, say, whether the computer has as much "computing power" as a human. Rather, trying to figure out whether the entity on the other end of the communication line is intelligent is carried out by probing intelligently. The interrogator is able to ask all kinds of questions, using his or her own sophisticated thinking, and carries out a dialog in an attempt to figure out whether the entity on the other end of the communication line understands things the way a human does. He or she then uses his or her own understanding (recognition) of the world in order to analyze the responses and assess the entity's authenticity.

What is the olfactory analogue of such a dialog, rich in human experience? An odor test in which a human tester is asked to compare real odors with their artificially produced versions is technical and fundamentally quantitative. It has little or nothing to do with the human experience of relating to the odors in question as part of their real world experience, and thus has nary a chance of capturing the elusive notion of recognizability. We are in the pre-Niépce-Bell period of olfactory reproduction, and want a way to convince ourselves that even first-generation ORSs produce whiffer output that is *recognizable*, not just technically similar to the original. Later on, perhaps far later on, when odor reproduction improves and becomes commonplace, we can devise quantitative ways that are analogous to the painstaking

---

capture scenes with low lighting, and a high-pitched violin might not have been recognizable on the receiving end of Bell's telephone. But these facts are beside the point: taking the technological constraints of the period into account, both techniques were doubtless able to adequately reproduce recognizable versions of all reasonable inputs, even then.

evaluation of modern photographic and sound equipment.[4] For now, we need to tap into the human's real-world experience, and it is in that respect that the naming issue arises in full force: verbally describing the arbitrary odors involved in a testing process is out of the question.

## 4. Immersion in Audio and Video

I would like to suggest a general method for testing recognizable odor reproduction, which involves a subtle combination of the human olfactory experience and a comparison of the real and the reproduced. The idea is to avoid the need to name or to verbally describe odors, by employing a multi-modal immersion approach, taking advantage of the fact that we already have excellent reproduction methods for sight and sound.

The test involves the candidate system for olfactory reproduction and two humans (actually, teams thereof). The first is a *challenger*, whose role is to challenge the ORS's claim to validity. The challenger can be thought of as representing users of the ORS, i.e., its eventual customers, for when it passes the test and is deemed successful. The second is an honest *tester*, who is willing to spend time on this, but who has no vested interest in the test's results either way. In the next section, a somewhat naïve test will be first described, followed by the more subtle recommended one. Both variants call for the challenger to provide the tester, repeatedly, with small sets of odor-emitting situations, or scenes, in the form of short clips recorded using a video camera (which includes audio) coupled with the ORS's sniffer; each say, 8-10 seconds long. Every testing session thus employs a set of audio-video recordings and recordings of the corresponding odor fingerprints.[5] The clips can be prepared in whatever locations the challenger fancies: a bustling marketplace, a damp cave, the lion's cage in a zoo, a grandparent's attic, the depths of a jungle, a hospital's operating room, or the set of a TV cooking show.

As we shall see in the next section, the testing itself involves several sessions, in which the testers are asked to make certain decisions regarding the video clips and the odors. When implementing the tests, the number of actual testing sessions carried out is important, as is the number of people constituting the challenger and tester teams. Also, care should be taken to devise means for preventing lucky guessing on the part of the testers, and for eliminating outliers. I will not get into these details here; they can be worked out in the standard ways used in setting up many kinds of experiments involving human response. Also, when talking about the tester having access to the video clips and the odors used in the testing process, I shall assume that technical provisions have been taken regarding the output devices (e.g., the audio-video projection and sniffer emission), which allow the tester to view or sniff any of these, as often as he or she wants, at any time during the testing session.

---

[4] True indistinguishability is not viable even for sight and sound. Despite tremendous advances over the years, people are still able to tell the difference between viewing an actual scene (with one eye, to eliminate the stereoscopic effect) and a photograph thereof, even a very high quality one; and the same applies to sound. Hence, some will claim that total indistinguishability *a la* Turing will never be truly attained for artificial renditions of any of our senses, and hence that it is not what we should be striving for here either.

[5] Short clips require only a single odor fingerprint, and not a dynamic series that would capture odors changing over time.

Our use of audio and video makes it possible to immerse a human in familiar sensory information for reference, in effect "placing" the tester where the odor was captured. The fact that there is no need for verbal characterization also helps reduce the effect of any relevant cultural differences that may exist: as long as the challenger chooses situations of which the testers can make real-world sense, it does not matter how a particular person perceives odors. The test must only verify that the whiffer-generated output adequately captures the original input odor, in a way that substantiates its recognizability in the video/audio setting by the human tester, regardless of how different people would have chosen to describe it, or even whether they could have done so at all.

## 5. The Test

Here is our first version of a test for the fidelity of an odor reproduction system. As we shall see soon, it is rather naïve.

> *Simple lineup*: The test involves several sessions. In each, the tester is given a small fixed number of challenger-produced video clips – say, between 5 and 10 – but is given the whiffer output of only one of them, without knowing which of the clips is its origin. The tester's role is to try to match the odor to the correct clip.

This simple lineup matching can be viewed as a straightforward human recognition task, a la Niépce and Bell, but, as discussed above, one that avoids the need to name or describe what is being smelled (or, for that matter, what is being seen and heard in the video clip); all the tester has to do is decide which of the video clips is most likely to have produced the odor he or she is given. Repeated success on a variety of sets of clips supplied by the challenger validates the ORS. Of course, the immersion idea is needed for testing only, to circumvent the naming and description problem for olfaction. Once validated, the ORS can be used for odor reproduction without the need for additional sense modalities.[6]

There is, however, a rather serious problem with this naïve test, at least in the way it has been set up. The challenger could be too eager to disqualify the ORS, producing sets of situations that are very much alike or ones that require special expertise on the part of an average tester, such as different glasses of wine with their labelled bottles nearby. Worse, the challenger may be downright vicious, recording, for example, a peaceful countryside, but with the video camera's back facing the opening of a damp cave. The on-location sniffer will capture the strong odor of the cave, which is nowhere to be seen in the clip, guaranteeing the test's failure. This can be partially alleviated by having the testers themselves play part of the role of the challenger, producing a large set of clips and sniffer readings (and making sure that the sniffer device is indeed placed adjacent to the video camera and in the "right" direction, or that the video makes appropriate horizontal and/or vertical sweeps), and with the challenger in each session merely presenting the tester with a specific subset of these as a lineup and a particular whiffer output to be matched. Still, there is something much cleaner and more convincing about a test in which the testers are neutral and uninvolved, and are challenged with situations about which they know nothing in advance. They must then use all their knowledge of the world, with its rich variety of sights, sounds and odors, in carrying out their task.

---

[6] Immersion was not needed by Niépce and Bell's contemporaries, since human "understanding" of sight and sound is very different from that of olfaction; even crude renditions of the former evoke perceptions of sufficiently recognizable and identifiable nature.

To eliminate these difficulties, the actual test I propose now is more subtle. It involves the challenger producing an additional channel of captured input for each situation, over and above the audio/video clip and the sniffer fingerprint recording. An actual odor sample is to be collected at each recorded location, in a way that enables future release. There are several viable techniques for doing this, although, in contrast to the ever-lasting digital nature of video recordings and sniffer fingerprints, current technology places limits on the duration for which the actual collected molecules are able to remain perceptually faithful to their capture time. Currently existing devices are based on microencapsulation or headspace technology [9, 16], but better techniques, with increased longevity of the samples, will probably surface in the future.

The proposed test is a sort of conditional, asymmetric lineup, which involves the original odor in addition to the artificially reproduced one.

> *Conditional lineup*: Here too, the test involves several sessions, in each of which the tester is given a 5-10 challenger-produced video clips and the odor corresponding to only one of them. However, there are now two testers (or teams thereof), where, unbeknown to them, the first is given the actual collected odor sample and the second is given the whiffer output corresponding to the same clip. The goal is for the second tester to succeed in the matching whenever the first one does. Sessions for which the first tester provides an incorrect match are ignored.

The conditional requirement, whereby we require success from the tester with the artificial odor only when the one with the original odor succeeds, appears to be a good way to balance the need for rigorous testing with the hard-to-define power of human odor perception. If humans are unable to "recognize" the real odor when immersed in its audio-video habitat, we cannot require them to be able to do so with the reproduced one. An ORS does its job well if humans are able to recognize the artificial whenever they can recognize the real thing.

The conditional lineup test is thus heavily inspired by the way humans recognize reproduced photos and audio, ever since the very early days of Niépce and Bell. However, it also indirectly checks the imitation facet of the ORS, in that the similarity of the real and the artificial (indirectly here, via the testing team's success in the identification task) are both considered, so that it is inspired also by Turing's idea of checking how well the artificial imitates the real. Moreover, the two main aforementioned difficulties are avoided: our test employs video and audio immersion to get around the naming and verbal description problem, and it eliminates unfair challenges by never requiring of the reproduced odor what we do not require of the original.

## 6. Discussion

Despite the advantages listed above, the proposed testing method has its weaknesses. One is that simple versions of the test, arranged by a less demanding challenger, might be able to reproduce only the very basic characteristics of an odor. An ORS will be able to pass such a non-stringent version even if it is only able to carry out the olfactory analogue of a sound system capturing only some part of the underlying tune from a symphonic orchestra playing a rich piece of music, or a camera that is able to make out only the rough, but identifiable, black-and-white outline of a figure. Thus, our test places a heavy responsibility on the challenger, who will have to be diligent enough in selecting the sight and sound scenes for the tester, so that reproducing, say, just the scent of plain bar of chocolate from the far more subtle odor of a rich chocolate-based dessert will in fact cause the test to fail.

The conditional immersion-based lineup proposed here is not the only test that comes to mind. Significant among alternative tests are those in which testers are requested to compare several odors, say, by singling out one that is different from the others, or those in which they are requested to determine which of several is "closest" in their opinion to the real one. As mentioned earlier, these are technical in nature, and have less to do with human's real world experience. It is somewhat akin to comparing wavelengths of pixels on a screen with those of the actual points in the original scene. I will not get into these possibilities further here, except to remark once again that the use of immersion here is in the interest of catering for the deeper issue of human recognizability in the context of the real world, with the odors embedded, so to speak, in their appropriate audio and video habitat.

One may also devise a series of tests with increasingly demanding challenges, which, rather than the ORS being labelled with just pass or no-pass, the result could be a grade of how well it does its job. Lumping together all manner of chocolate would be on the low end of the spectrum, good enough as a first stab at the problem, Niepce-Bell-like, whereas being able to distinguish the odor of two very similar chocolate-based cakes, such as a Viennese Sachertorte and its Demel bakery variant, would be on its high end.

Another weakness of the conditional lineup test stems from its most novel facet – the immersion idea. Many odors are not naturally associated with specific sight-sound scenes, and, dually, there could be numerous odors "legitimately" associated with a given scene. Such situations can often be dealt with adequately by challengers who make sure to produce clips that are sufficiently informative, and engaging more-capable-than-average testers. Faithful reproduction of wine odors, for example, could be tested by indeed videoing the wine bottles with their labels, perhaps also showing the wine rolling around in the glass or held against the light, and employing wine experts as the testers; and similarly for haute cuisine or haute pâtisserie. One thing, however, does seem certain: the proposed test will work well for what one might call *scene-related odors*, and these, I claim, constitute the vast majority of odors a typical human will usually encounter.[7]

Finally, the immersion idea could do with a better understanding of the interactions between olfaction and the senses of sight and sound, both from a psychological point of the view and from an epistemic one. The strong connections between olfaction and taste are well known, as are interesting phenomena regarding the two, such as 'phantom aromas' used by the food industry, and the much studied notion of odor-induced taste. Are there similar phenomena involving olfaction and vision? A positive answer would clearly be relevant to the test proposed here. As one of the reviewers of this article put it, if the video clip of a sizzling steak caused one to identify a whole family of unrelated smells with 'steak', then the inability to distinguish the odor of a real steak from the ORS-generated one would not tell us much. I could not agree more. This is definitely an exciting area for future research, which should also address the question of the extent to which our idea of immersion in video and audio might actually help in identifying odors that would otherwise not be adequately recognizable: thus an ORS that passes even a stringent version of our test may not perform as well later, when stripped of the helpful immersion.

---

[7] Nevertheless, non-scene-related cases can often be dealt with too. An unusual chemical, not normally associated with any conventional sight and sound scene, can be tested reasonably well by a challenger who uses a set of scenes consisting of nine "normal" ones and the tenth being an open laboratory bottle. A diligent tester will correctly match the chemical with the latter by elimination. Alternatively, we can have the challenger show only normal scenes, but allow "none of the above" answers too.

## 6. Conclusion

Full odor reproduction systems, which deal adequately with any input odor, might be long in coming, but I believe we will begin to see initial attempts quite soon. Just as technologies for the reproduction of sight and sound have changed and improved radically since the pioneering work of Niépce and Bell, so will future years see radically new ways of capturing, communicating and reproducing odor in ways recognizable by humans. This paper proposes a criterion for assessing the quality of such systems, in the form of a test for the human-centric adequacy of such systems when they do arrive, which, I think, is important in its own right. An (imaginary) ORS that somehow produces replicas of input odors that chemically trigger close-to-identical human perception will clearly pass our test with flying colors (or perhaps, stretching the linguistic metaphor perhaps a bit too much, one should say 'with flying odors'…).

I am hopeful that this paper will trigger further thinking about the extremely difficult, but exciting problem of achieving satisfactory artificial olfactory reproduction, hand in hand with developing the best methods for testing the solutions.

**Acknowledgments**

Special thanks go to Assaf Marron for providing numerous detailed, thoughtful and constructive comments on all parts of the paper, in all its writing stages. I am grateful to Hadas Lapid for illuminating discussions on a preliminary version, and to Yair Harel for his artful rendition of the human sniffer (bottom-right in Fig. 4). The three reviewers, who appear to have come from quite different fields of research, provided extremely helpful comments. In particular, two of the reviews triggered the last paragraph of the Discussion section. Parts of the writing were carried out during a visiting position in Wil van der Aalst's group at the Technical University of Eindhoven.